\documentclass[12pt]{article}  
\usepackage{graphicx}
\def \bea{\begin{eqnarray}}
\def \beq{\begin{equation}}
\def \eeq{\end{equation}}
\def \esix{E$_{\rm 6}$}

\begin{document}

\centerline{\bf STERILE NEUTRINOS IN THE GRAND UNIFIED GROUP E$_6$}
\medskip
\centerline{\it Jonathan L. Rosner}
\medskip
\centerline{\it Enrico Fermi Institute and Department of Physics}
\centerline{\it University of Chicago, Chicago, IL 60637}
\bigskip

Recent short-baseline neutrino oscillation experiments have led us to ask
whether an adequate description of neutrino masses and mixings can be
obtained using just three neutrinos which are doublets of SU(2)$_L$
[left-handed electroweak SU(2)] \cite{sbl}.  One or more light ``sterile''
[SU(2)$_L$-singlet] neutrinos may be required in addition.  This short note
is to remind readers of the opportunity for three light sterile neutrinos
within the grand unified group \cite{GRS} E$_6$.  Some details may be found
in Ref.\ \cite{cmts}.

Ths standard model group SU(3)$_{\rm color} \times$ SU(2)$_{\rm L} \times$
U(1) can be incorporated into a grand unified group.  Popular
candidates include SU(5) $\subset$ SO(10) $\subset$ \esix.
Each quark and lepton family constitute a $5^* + 10$ of SU(5) (without
right-handed neutrinos).  Adding a right-handed neutrino $N$ [an SU(5)
singlet] to each such hypermultiplet, one gets a 16 of SO(10).  A
right-handed neutrino can pair with a left-handed one to
generate a Dirac mass $m_D$ as occurs for charged leptons and quarks.  However,
the neutrality of the right-handed neutrino under the standard model group
allows it to have a large Majorana mass $M$, leading via the seesaw mechanism
to light-neutrino masses $m = m_D^2/M$.  At this stage there are three
light neutrinos (mostly electroweak doublets) and three heavy ones (mostly
electroweak singlets).

Proceeding beyond SO(10), a 10-plet of that group [a 5 + $5^*$ of SU(5)] can
be added to each quark and lepton family.  It consists of quarks which are
singlets under SU(2)$_L$ and SU(2)$_R$, and leptons which are doublets under
both.  To form the smallest \esix~representation, a 27-plet, all one need
add is another singlet $n$ of SO(10).  The $n$ is a sterile neutrino with
neither L nor R isospin.  As there is nothing left for it to pair up with,
it is naturally light and of Majorana nature.  It would have to mix, however,
with standard model light neutrinos in order to account for some of the
reported anomalies in short-baseline neutrino experiments.

If grand unified groups are the source of sterile neutrinos (either 
$N$ or $n$) then they come in threes.  While present data do not show a
significant advantage in enlarging the number of sterile neutrinos to three,
this possibility should be kept in mind when searching for motivations for
such neutrinos.


\begin{thebibliography}{99}

\bibitem{sbl} See, for example,
S.~J.~Brice, S.~Geer, D.~Harris, B.~Kayser, S.~Parke, C.~Polly, R.~Tschirhart,
G.~Zeller {\it et al.},
``Short-Baseline Neutrino Focus Group Report,'' Fermilab report No.\
FERMILAB-FN-0947;
J.~M.~Conrad, C.~M.~Ignarra, G.~Karagiorgi, M.~H.~Shaevitz and J.~Spitz,
``Sterile Neutrino Fits to Short Baseline Neutrino Oscillation Measurements,''
arXiv:1207.4765 [hep-ex].

\bibitem{GRS} F. G\"ursey, P. Ramond, and P. Sikivie, Phys.\ Lett.\ {\bf B60},
177 (1976).

\bibitem{cmts} J. L. Rosner, ``E$_6$ and Exotic Fermions,'' Comments on Nucl.\
and Particle Phys.\ {\bf 15}, 195 (1986).
\end{thebibliography}
\end{document}